\documentclass[a4paper,12pt]{article}
\usepackage{amsmath,amsfonts,amssymb}
\usepackage[mathscr]{eucal}
\usepackage{natbib}
\newtheorem{lemma}{Lemma}
\newtheorem{definition}[lemma]{Definition}
\newtheorem{theorem}[lemma]{Theorem}
\newtheorem{conjecture}[lemma]{Conjecture}

\newtheorem{corollary}[lemma]{Corollary}

\newtheorem{postulate}[lemma]{Postulate}
\newtheorem{fact}[lemma]{Fact}

\newcommand{\qed}{\bf Q.E.D.} 
\newenvironment{proof}{{\bf Proof}: }{\;\;\qed}

\newcommand{\reals}{{\mathbb R}}

\begin{document}
\title{On Minkowskian Branching Structures} \author{Leszek Wro{\'
    n}ski and Tomasz Placek%
  \thanks{We read earlier versions of this paper at the seminar `Chaos
    and Quantum Information' held at the Jagiellonian University in
    Krak{\' o}w on April 16, 2007 and at the seminar `On Determinism'
    held at the University of Bonn on April 20, 2007. For comments and
    valuable discussions we are grateful to the audiences, and in
    particular, to Dr. Thomas M{\" uller}. Authors' address:
    Department of Philosophy, Jagiellonian University, Grodzka 52,
    31-044 Krak{\' o}w, Poland;\; LW's email: {\tt elwro1@gmail.com}
    and TP's email: {\tt uzplacek@cyf-kr.edu.pl}.%
}}
\date{\today}
\maketitle
\begin{abstract}
  Contrary to its initial idea, Belnap's (\citeyear{belnap92}) theory
  of Branching Space-Times (BST) has models in which histories do not
  resemble relativistic space-times or any other physical space-times.
  The aim of this paper is to define a certain class of BST models,
  called "Minkowskian Branching Structures" (MBS), in which histories
  are isomorphic to Minkowski space-time. By focusing on these models
  rather than on general BST models, we hope that one may be able to
  improve on earlier BST analyzes of physical phenomena.  Also,
  introducing MBS' sets the stage for recent discussions about whether
  or not `branching is a bad idea', physically speaking.
\end{abstract}

\section{Introduction} 
Branching Space-Times (BST) of Belnap (\citeyear{belnap92}) is a
possible-worlds theory that intends to combine, in a rigorous fashion,
indeterminism, relativity and modalities. Yet, ironically, BST allows
for histories that resemble neither relativistic space-times nor any
other physical space-times. M{\"u}ller, Belnap and Kishida
(\citeyear{muller06:_funny}) discovered a structure called $M_2$
which, while being a BST model, is in no way similar to the intended
pictures of the theory.  Our aim in this paper is thus to define a
certain class of BST models called "Minkowskian Branching Structures"
(MBS), in which histories are isomorphic to Minkowski space-time. In
defining the notion we follow the lead of \citet{muller_nato}, yet
with two crucial diversions.  First, we remove M{\"u}ller's finiteness
assumptions, as they prohibit the introduction of `interesting'
infinite structures. Second, we improve on M{\"u}ller's failed proof
of the most desired feature of MBS, namely that every history is
isomorphic to Minkowski space-time.  To this end we assume a certain
topological postulate.

By concentrating on MBS' rather than general BST models, one may avoid
a certain awkwardness inherent in earlier analyses of some physical
phenomena, done from a BST perspective.\footnote{To give some
examples, the Bell / Aspect Theorems were analyzed in
\citet{belnap-szabo}, \citet{kowalski99}, and \citet{mueller01}.  A
branching perspective on quantum holism was put forward in
\citet{placek04_holism}. A BST reading of the consistent histories
interpretation of quantum mechanics was given in \citet{mueller07}.
Some questions from the philosophy of time were discussed, from a BST
perspective, in \citet{mueller_present}.} In essence, they tried to fit
information about the location of objects into models with no metric
features. This problem will not arise in our structures.

Singling out the class of MBS models has yet another advantage: It has
been alleged in the philosophy of physics community that branching is
a bad idea.\footnote{See for instance J. Earman's talk at the Ontology
  of Space-Time conference, Montreal 2006.} Yet, to sensibly discuss
the pros and cons of branching, as applied to physics, one had better
have a class of BST models which have some pretense at being
`physically realistic'.

The paper is organized as follows. In Section~\ref{sec1} we review
some definitions and theorems of BST which we will need later.
Section~\ref{sec2} defines and discusses Minkowskian Branching
Structures. The final section~\ref{concl} states our conclusions and
an open problem.

\section{Branching Space-Times}{\label{BST}}\label{sec1}
The theory of Branching Space-Times (BST) of Belnap
(\citeyear{belnap92}), combines objective indeterminism and relativity
in a rigorous way. Its primitives are a nonempty set $W$ (called ``Our
World'', interpreted as the set of all possible point events) and a
partial ordering $\leqslant$ on $W$, interpreted as a ``causal order''
between point events.

There are no ``Possible Worlds'' in this theory; there is only one
world, Our World, containing all that is (timelessly) possible.
Instead, a notion of ``history'' is used, as defined below:

\begin{definition}{\label{updir}} A set $h \subseteq W$ is
  upward-directed iff $\forall e_1, e_2 \in h \; \exists e \in h$ such
  that $e_1 \leqslant e$ and $e_2 \leqslant e$.

  A set $h$ is maximal with respect to the above property iff $\forall
  g \in W$ such that $g \varsupsetneq h$, $g$ is not upward-directed.

  A subset $h$ of $W$ is a history iff it is a maximal upward-directed
  set.

For histories $h_1$ and $h_2$, any maximal element in $h_1 \cap h_2$
is called a choice point for $h_1$ and $h_2$.
\end{definition}

A very important feature of BST is that histories are closed downward:
if $e_1 \leqslant e_2$ and $e_1 \notin h$, then $e_2 \notin h$. In
other words, there is no backward branching among histories in BST. No
two incompatible events are in the past of any event; equivalently:
the past of any event is ``fixed'', containing only compatible events.

We will now give the definition of a BST model; for more information about BST in general see \cite{belnap92}.

\begin{definition}{\label{BSTmodel}}
  $\langle W, \leqslant \rangle$ where $W$ is a nonempty set and
  $\leqslant$ is a partial ordering on $W$ is a model of BST if and
  only if it meets the following requirements:
\begin{enumerate}
\item The ordering $\leqslant$ is dense.
\item $\leqslant$ has no maximal elements.
\item Every lower bounded chain in $W$ has an infimum in $W$.
\item Every upper bounded chain in $W$ has a supremum in every
history that contains it.
\item (Prior choice principle ('PCP')) For any lower bounded chain $O \in
h_1 - h_2$ there exists a point $e \in W$ such that $e$ is maximal in $h_1 \cap h_2$ and $\forall e^{'} \in O \;
e < e^{'}$.
\end{enumerate}

\end{definition}

Note that BST has the causal ordering $\leqslant$, yet in general it
has neither space-time points nor a space-time (i.e., a collection of
space-time points). The intuition behind space-time points is that we
want to be able to say that something happens ``in the same space-time
point'' in different histories. A triple $\langle W, \leqslant, s
\rangle$ is a ``BST model with space-time points'' (BST+S) iff
$\langle W, \leqslant \rangle$ is a BST model and $s$ (from the
expression ``space-time point'') is an equivalence relation on $W$
such that 1) for each history $h$ in $W$ and for each equivalence
class $s(x)$, $x \in W$, the intersection $h \cap s(x)$ contains
exactly one element and 2) $s$ respects the ordering: for equivalence
classes $s(x), s(y)$ and histories $h_1, h_2$, $s(x) \cap h_1 = s(y)
\cap h_1$ iff $s(x) \cap h_2 = s(y) \cap h_2$, and the same for
``$<$'' and ``$>$''. As \citet{mueller05} shows, not every BST model
can be extended to a BST+S model. Moreover, there are BST+S models,
whose space-time $S$ has nothing in common with any physically
motivated space-time -- the structure $M_2$ of M{\" u}ller, Belnap and
Kishida (\citeyear{muller06:_funny}) is an example.

\section{Introducing Minkowskian Branching Structures}{\label{MBS}}\label{sec2}
We are now going to single out a class of BST models in which
histories occur in Minkowski space-time. This part of our work is
based on M\"{u}ller's (\citeyear{muller_nato}) theory.

\noindent
The points of  Minkowski space-time are elements of $\reals^4$,
e.g. $x\! =\! \langle x^0, x^1, x^2, x^3 \rangle$, where the first element
of the quadruple is the time coordinate. The Minkowski space-time
distance is a function $D^2_M : \reals^4 \times \reals^4 \rightarrow
\reals$ defined as follows (for $x,y \in \reals^4$):
\begin{equation}
D^2_M(x,y) := -(x^0 - y^0)^2 + \sum_{i=1}^3(x^i - y^i)^2
\end{equation}
The natural ordering on the Minkowski space-time, call it
``Minkowskian ordering $\leqslant_M$'', is defined as follows ($x,y
\in \reals^4$):
\begin{equation}
  x \leqslant_M y \mbox{ iff } D^2_M(x,y) \leqslant 0 \mbox{ and } x^0 \leqslant y^0
\end{equation}

We will say that two points $x, y \in \reals^4$ are space-like related
(``SLR'' for short) iff neither $x \leqslant_M y$ nor $y \leqslant_M
x$. Naturally, $x <_M y$ iff $x \neq y$ and $x \leqslant_M y$.

Now we need to provide a framework for ``different ways in which
things can happen'' and for filling the space-times with content. For
the first task we will need a set $\Sigma$ of labels $\sigma, \eta,
...$. (In contrast to \cite{muller_nato}, we allow for any cardinality
of $\Sigma$). For the second task, we will use a so-called ``state''
function $S : \Sigma \times \reals^4 \rightarrow P$, where $P$ is a
set of point properties (on this we just quote M\"{u}ller saying
``finding out what the right $P$ is is a question of physics, not one
of conceptual analysis'').

One could ask about the reasons for an extra notion of a ``scenario''.
Why don't we build histories just out of points from $\reals^4 \times
P$? The reason is that a member of BST's Our World has a fixed past.
If two different trains of events lead to exactly the same event $E
\in \reals^4 \times P$, the situation gives rise to two different
point events, two different members of $W$. In contrast: for a point
$\langle x, p_0 \rangle$ from $\reals^4 \times P$ there can exist two
different points $\langle y, p_1 \rangle$ and $\langle y, p_2 \rangle$
from $\reals^4 \times P$ such that $y <_M x$. This would be a case of
backward branching, so the set $\reals^4 \times P$ is not a good
candidate for the master set $W$ of any BST model.

The idea behind the concept of scenario is that every scenario
corresponds to an $\reals^4$ space filled with content\footnote{Fix a
  scenario $\alpha$. The above mentioned corresponding space filled
  with content is $A \subseteq \reals^4 \times P$ such that $\langle
  x, p \rangle \in A$ iff $S(\alpha,x)=p$.}, where the content derives
from the elements of $P$. Assuming that a certain state function $S$
is given, for any $\sigma, \eta \in \Sigma$ the set $C_{\sigma\eta}
\subset \reals^4$ is the set of ``splitting points'' between scenarios
$\sigma$ and $\eta$; intuitively it is the set of points in which a choice
between the two scenarios is made. All members of $C_{\sigma\eta}$
have to be space-like related. Of course a choice between $\sigma$ and
$\eta$ is a choice between $\eta$ and $\sigma$, so $C_{\sigma\eta} =
C_{\eta\sigma}$. The BST axiom of prior choice principle motivates our
postulate that any two different scenarios split. Formally: $\forall
\sigma, \eta \in \Sigma \; (\sigma \neq \eta \Rightarrow
C_{\sigma\eta} \neq \emptyset)$.

The next requirement concerns triples of scenarios. Any set
$C_{\sigma\eta}$ determines a region in which both scenarios coincide:
namely, that part of $\reals^4$ that is not in the Minkowskian sense
strictly above any point from $C_{\sigma\eta}$. Following M\"{u}ller
we call it the region of overlap $R_{\sigma\eta}$ between scenarios
$\sigma, \eta$ defined as below:
\begin{equation}\label{overlap}
  R_{\sigma\eta} := \{ x \in \reals^4 | \neg \exists y \in C_{\sigma\eta} \; y <_M x \}
\end{equation}
(Of course it follows that for any $\sigma, \eta \in \Sigma
C_{\sigma\eta} \subseteq R_{\sigma\eta}$.) Assuming the sets
$C_{\sigma\eta}$ and $C_{\eta\gamma}$ are given, we get two regions of
overlaps $R_{\sigma\eta}$ and $R_{\eta\gamma}$. At the points in the
intersection of those two regions $\sigma$ coincides with $\eta$ and
$\eta$ coincides with $\gamma$, therefore by transitivity $\sigma$
coincides with $\gamma$. In general we can say that for any $\sigma,
\eta, \gamma \in \reals^4$
\begin{equation}\label{overlap2}
R_{\sigma\gamma} \supseteq R_{\sigma\eta} \cap R_{\eta\gamma}
\end{equation}
which, translated to a requirement on sets of splitting points, is
\begin{equation}\label{overlap3}
\forall x \in C_{\sigma\gamma} \exists y \in C_{\sigma\eta} \cup C_{\eta\gamma} y \leqslant_M x.
\end{equation}

In his paper M\"{u}ller put another requirement on $C_{\sigma\eta}$:
finitude. The motivation was to exclude splitting along a
``simultaneity slice''. The strong requirement of finitude, however,
excludes many more types of situations in which splitting is not
continuous or happens in a region of space-time of  finite diameter.
In the present paper we drop this requirement, not putting any
restrictions on the cardinality of $C_{\sigma\eta}$ for any $\sigma,
\eta \in \Sigma$. As a side-note, this leads to the fact that in some
models there may be choice points which are not intuitively connected
with any splitting point. For details, see \ref{spcp} in the Appendix.

The state function assigns to each pair $\langle \mbox{a label from
  $\Sigma$, a point from $\reals^4$} \rangle$ an element of $P$.
Colloquially, the state function tells us what happens at a certain
point of the space-time in a given scenario.\footnote{We can look at
  the situation from a slightly different perspective: every label
  $\sigma$ is assigned a mapping $S_\sigma$ from $\reals^4$ to $P$;
  see also the previous footnote.}

After \citet{muller_nato}, we now proceed to construct the elements of
the MBS version of Our World; they will be equivalence classes of a
certain relation $\leqslant_S$ on $\Sigma \times \reals^4$. For
convenience, we write the elements of $\Sigma \times \reals^4$ as
$x_\sigma$, where $x \in \reals^4, \sigma \in \Sigma$. The idea is to
``glue together'' points in regions of overlap; hence the relation is
defined as below:
\begin{equation}\label{S1}
x_\sigma \equiv_S y_\eta \mbox{ iff } x = y \mbox{ and } x \in R_{\sigma\eta}.
\end{equation}
M\"{u}ller provides a simple proof of the fact that $\equiv_S$ is an
equivalence relation on $\Sigma \times \reals^4$; therefore, we can
produce a quotient structure. The result is the set $B$, which is the
MBS version of Our World:
\begin{equation}\label{B1}
B := (\Sigma \times \reals^4) / \equiv_S \; \; = \{ [x_\sigma] | \sigma \in \Sigma, x \in \reals^4 \},
\end{equation}
where $[x_\sigma]$ is the equivalence class of $x$ with respect to the
relation $\equiv_S$:
\begin{equation}\label{B2} [x_\sigma] = \{ x_\eta | x_\sigma \equiv_S
  x_\eta \}.
\end{equation}
Next, we define a relation $\leqslant_S$ on $B$:
\begin{equation}\label{SS1}
[x_\sigma] \leqslant_S [y_\eta] \mbox{ iff } x \leqslant_M y \mbox{ and } x_\sigma \equiv_S x_\eta ,
\end{equation}
which (as M\"{u}ller shows) is a partial ordering on $B$.

The goal would now be to prove that $\langle B, \leqslant_S \rangle$
is a model of BST. To do so, and in particular to prove the prior
choice principle and requirement no. 4 from definition \ref{BSTmodel},
we need to know more about the shape of the histories in MBS - that
they are the intended ones.

\subsection{The shape of MBS histories}\label{shape} We would like
histories, that is, maximal upward-directed sets, to be sets of
equivalence classes $[x_\sigma]$ (with respect to $\equiv_S$) for $x
\in \reals^4$ for some $\sigma \in \Sigma$. In other words, we wish to
unambiguously refer to any history by a label from $\Sigma$, requiring
one-to-one correspondence of the sets of histories and labels. This is
M{\"u}ller's (\citeyear{muller_nato}) Lemma 3 and our

\begin{theorem}{\label{th5}} Every history in a given MBS is of the
  form $h = \{ [x_\sigma]|x \in \reals^4 \}$ for some $\sigma \in
  \Sigma$.
\end{theorem}
The problem is that, aside from minor brushing up required by the
proof of the ``right'' direction, the proof of the ``left'' direction
supplied in \cite{muller_nato} needs to be fixed, as it does not
provide adequate reasons for nonemptiness of an essential intersection
$\bigcap \Sigma_h (z_i)$. More on that below. Let us divide the above
theorem into two lemmas (\ref{l321} and \ref{l322}), corresponding to
the directions and prove the ``right'' direction first. Until we prove
the theorem we refrain from using the term ``history'' and substitute
it with a ``maximal upward-directed set'' for clarity.

\begin{lemma}{\label{l321}} If $h = \{ [x_\sigma]|x \in \reals^4 \}$
  for some $\sigma \in \Sigma$, then $h$ is a maximal upward-directed
  subset of $B$.
\end{lemma}

\begin{proof}
  Let us consider $e_1, e_2 \in h,e_1 = [x_\sigma], e_2 = [y_\sigma]$.
  Since $x,y \in \reals^4$, there exists a $z \in \reals^4$ such that
  $x \leqslant_M z$ and $y \leqslant_M z$. Therefore, $[x_\sigma]
  \leqslant_S [z_\sigma]$ and $[y_\sigma] \leqslant_S [z_\sigma]$, and
  so $h$ is upward-directed.

  For maximality, consider a $g \subseteq B, g \varsupsetneq h$ and
  assume $g$ is upward-directed. It follows that there exists a point
  $[x_\eta] \in g-h$ such that $[x_\eta] \neq [x_\sigma] \in h$. Since
  both points belong to $g$, which is upward-directed, there exists
  $[z_\alpha] \in g$ (note that we are not allowed to choose $\sigma$
  as the index at that point) such that $[x_\eta] \leqslant_S
  [z_\alpha]$ and $[x_\sigma] \leqslant_S [z_\alpha]$.  Therefore,
  $x_\eta \equiv_S x_\alpha \equiv_S x_\sigma$, and so we arrive at a
  contradiction by concluding that $[x_\eta] = [x_\sigma]$.
\end{proof}

The proof of the other direction is more complex and, what might be
surprising, involves a topological postulate. First, we will need a
simple definition:

\begin{definition}{\label{Sigma}}
  For a given maximal upward-directed set $h$ and a point $x \in
  \reals^4$, $\Sigma_h(x) := \{ \sigma \in \Sigma | [x_\sigma] \in h
  \}$.
\end{definition}

Consider now a given maximal upward-directed set $h \subseteq B$. With
every lower bounded chain $L \subset \reals^4$ we would like to
associate a topology (called a ``chain topology'') on the set of
$\Sigma_h(\inf(L))$.  We define the topology by describing the whole
family of closed sets, which is equal to $\{ \emptyset,
\Sigma_h(\inf(L)) \} \cup \{ \Sigma_h(l) | l \in L \} \cup \{ \cap \{
\Sigma_h(l) | l \in L \} \}$. (Because $L$ is a chain, it is evident
that the family is closed with respect to intersection and finite
union). The postulate runs as follows:

\begin{postulate}{\label{p1}}
  For every maximal upward-directed set $h \subseteq B$ and for every
  lower bounded chain $L \subset \reals^4$ the ``chain topology''
  described above is compact.
\end{postulate}

It is easily verifiable that in such a topology $\{ \Sigma_h(l) | l \in L \}$ is a centred family of closed sets
(every finite subset of it has a nonempty intersection). It is provable that the above postulate is equivalent
to the following corollary:

\begin{corollary}{\label{c1}}
  For every maximal upward-directed set $h \subseteq B$ and for every
  chain $L \subset \reals^4$, $\empty \bigcap \{ \Sigma_h(l) | l \in L
  \} \neq \emptyset$.
\end{corollary}

\begin{lemma}{\label{l322}}
  If $h$ is a maximal upward-directed subset of $B$, then $h = \{
  [x_\sigma]|x \in \reals^4 \}$ for some $\sigma \in \Sigma$.
\end{lemma}

The structure of the proof mimics M\"{u}ller's
(\citeyear{muller_nato}) proof. It is divided into three parts, the
first and the last being reproduced here. On the other hand, the
second part contains an error (as stated above, the statement that
$\bigcap \Sigma_h (z_i) \neq \emptyset$ is not properly justified) and
relies on the assumption that for every history $h$ and point $x \in
\reals^4$ the set $\Sigma_h(x)$ is at most countably infinite. We wish
both to drop this assumption and correct the proof using the above
topological postulate.

\begin{proof}
  Suppose that $h$ is a maximal upward-directed subset of $B$. In
  order to prove the lemma, we will prove the following three steps:

  1. If for some $\sigma, \eta \in \Sigma$ both $[x_\sigma] \in h$ and
  $[x_\eta] \in h$, then $x_\sigma \equiv_S x_\eta$.

2. There is a $\sigma \in \Sigma$ such that for every $\eta$, if $[x_\eta] \in h$, then $x_\eta \equiv_S
x_\sigma$.

3. With the $\sigma$ from step 2, $h = \{ [x_\sigma] | x \in \reals^4 \}$.

\medskip

\textbf{Ad. 1.} Since $h$ is maximal by assumption, there exists a
$[y_\gamma] \in h$ such that $[x_\sigma] \leqslant_S [y_\gamma]$ and
$[x_\eta] \leqslant_S [y_\gamma]$. These last two facts imply that
$x_\sigma \equiv_S x_\gamma \equiv_S x_\eta$, so by transitivity of
$\equiv_S$ we get $x_\sigma \equiv_S x_\eta$.

\textbf{Ad. 2.} Assume the contrary: $\forall \sigma \in \Sigma \;
\exists [x_\eta] \in h, x_\eta \not\equiv_S x_\sigma$.

Take a point $[y_\kappa] \in h$. Accordingly, $\Sigma_h(y) \neq
\emptyset$.

For each scenario $\sigma_\alpha \in \Sigma_h(y)$ we define a set
$\Theta_\alpha = \{ x \in \reals^4 \; | \; \exists \eta \in
\Sigma_h(y): \; [x_\eta] \in h \; \wedge \; x_{\sigma_\alpha}
\not\equiv_S x_\eta \}$, which by our assumption is never empty.
Colloquially, it is a set of the points that make the scenario a wrong
candidate for the proper scenario from our lemma - the scenario
``doesn't fit'' the history at those points. For each scenario
$\sigma_\alpha$ we would like to choose a single element of
$\Theta_\alpha$, and to that end we employ a choice function $T$
defined on the set of subsets of $\reals^4$ such that
$T(\Theta_\alpha) \in \Theta_\alpha$, naming the element chosen by it
as follows: $T(\Theta_\alpha) := x_\alpha$.\footnote{Bear in mind that
  since $\alpha$ is a number serving just as an index for scenarios,
  $x_\alpha$ (like $x_\beta$ in the line below inequality
  \ref{contr1}) is a point from $\reals^4$ and does not denote a point
  - scenario pair.}

Observe that we will arrive at a contradiction if we prove that
\begin{equation}{\label{contr1}}
  \bigcap_{\sigma_\alpha \in \Sigma_h(y)} \Sigma_h(x_\alpha) \neq \emptyset
\end{equation}
(since for any $\sigma_\beta \in \Sigma_h(y) \; \sigma_\beta \notin
\Sigma_h(x_\beta)$). In order to apply our topological postulate, we
will construct a chain $L = \{ z_0, z_1,$ ... ,$ z_\omega, ... \}$ of
points in $\reals^4$. It will be lower bounded by its initial element
$z_0$. Moreover, we want it to be vertical, since in this way it will (if
it does not have an upper bound itself) contain an upper bound of any
point in $\reals^4$, which will be needed in our proof.

We first define a function ``$up$'' which, given two points $a, b \in
\reals^4$, will produce a point $c \in \reals^4$ such that $c$ has the
same spatial coordinates as $a$ but is above $b$. In other words, if
$a = \langle a^0, a^1, a^2, a^3 \rangle \in \reals^4$, $b = \langle
b^0, b^1, b^2, b^3 \rangle \in \reals^4$, $up(a,b) := \langle
a^0+(\sum\limits_1^3 (a^i - b^i)^2)^{1/2}, \; a^1, a^2, a^3 \rangle
\in \reals^4$. Notice that $up$ is not commutative.

We proceed to define the above mentioned chain $L$ in the following way:

\emph{1}.$z_0 = up(y,x_0)$.

$z_1 = up(z_0,x_1)$.

Generally, $z_{k+1} \; = \; up(z_k, x_{k+1})$.

\emph{2}. Suppose $\rho$ is a limit number. Define $A_\rho := \{
z_\beta \; | \; \beta < \rho \}$\footnote{Again, $\beta$ is just an
  index, not a scenario, so $A_\rho$ is a subset of $\reals^4$.}. As
you can see, $A_\rho$ is the part of our chain we have managed to
construct so far. We need to distinguish two cases:

a) $A_\rho$ is upper bounded with respect to $\leqslant_M$. Then it
has to have ``vertical'' upper bounds $t_0, t_1 ...$ with spatial
coordinates $t^i_n = z_0^i$ $(i = 1,2,3)$. In this case, we employ the
above defined function $T$ to choose one of the upper bounds of
$A_\rho$:
\begin{equation}
  t_\rho := T( \{ t \in \reals^4 \; | \; \forall \beta < \rho \; z_\beta \leqslant_M t \wedge t^i = z_0^i (i =
  1,2,3)\}).
\end{equation}
Then we put $z_\rho := up(t_\rho, x_\rho)$, arriving at the next
element of our chain $L$.

b) if $A_\rho$ is not upper bounded with respect to $\leqslant_M$, then no matter which point in $\reals^4$ we
choose, it is possible to find a point from $A_\rho$ above it (since $A_\rho$ is vertical). Therefore, the set
\begin{equation}
  B_\rho = \{ t \in A_\rho \; | \; x_\rho \leqslant_M t \}
\end{equation}
is not empty. We put $[z_\rho] := T(B_\rho)$, arriving at the next element of our chain $L$.

Notice that in our chain it might happen that while $\alpha < \beta$,
$z_\beta \leqslant_M z_\alpha$, $z_0$ is a lower bound of $L$.
Therefore, our postulate \ref{p1} applies. By employing it and
corollary \ref{c1} we infer that
\begin{equation}\label{zety}
  \bigcap_{\sigma_\alpha \in \Sigma_h(y)} \{ \Sigma_h(z_\alpha) | z_\alpha \in L \} \neq \emptyset .
\end{equation}
By our construction of the chain $L$, for all $\alpha$ it is true that
$x_\alpha \leqslant_M z_\alpha$.  Therefore, $\Sigma_h(z_\alpha)
\subseteq \Sigma_h(x_\alpha)$. Thus, from \ref{zety} we immediately
get
\begin{equation}
  \bigcap_{\sigma_\alpha \in \Sigma_h(y)} \Sigma_h(x_\alpha) \neq \emptyset,
\end{equation}
which is the equation \ref{contr1} that we tried to show. Therefore, we
arrive at a contradiction and part 2 of the proof is complete.

\textbf{Ad. 3.} We have shown that there is a scenario $\sigma \in
\Sigma$ such that all members of $h$ can be identified as $[x_\sigma]$
for some $x \in \reals^4$. What remains is to show that the history
cannot ``exclude'' some regions of $\{ \sigma \} \times \reals^4$,
that is, to prove that for all $x \in \reals^4, [x_\sigma] \in h$. But
in lemma \ref{l321} we have shown that $\{ [x_\sigma] | x \in \reals^4
\}$ is a maximal upward-directed subset of $B$, so any proper subset
of it cannot be maximal upward-directed.
\end{proof}

By showing lemmas \ref{l321} and \ref{l322} we have proved theorem
\ref{th5}.

\subsection{The importance of the topological postulate}
So far it might seem that our topological postulate \ref{p1} is just a
handy trick for proving  lemma \ref{l322}. To show its importance
we will now prove that its falsity leads to the falsity of the lemma,
and then present an example of a structure in which the lemma does not
hold.

\begin{theorem}
  If  postulate \ref{p1} is false, then lemma \ref{l322} is also
  false.
\end{theorem}

\begin{proof}
  Assume that our topological postulate does not hold. Therefore, there
  exists a maximal upward-directed set $h \subseteq B$ and a lower
  bounded chain $L \subset \reals^4$ such that the chain topology is
  not compact. By rules of topology this is  equivalent to the fact
  that there is a centred family of closed sets with an empty
  intersection. But all closed sets in the chain topology form a chain
  with respect to inclusion. Of course, if a part of a chain has an
  empty intersection, a superset of the part also has an empty
  intersection. We infer that
\begin{equation}
\bigcap_{x \in L} \Sigma_h(x) = \emptyset ,
\end{equation}
from which, by definition \ref{Sigma}, we get that
\begin{equation}
\neg \exists \sigma \in \Sigma : \forall x \in L \; [x_\sigma] \in h ,
\end{equation}
so there is no scenario $\sigma$ such that $h = \{ [x_\sigma] | x \in \reals^4 \}$. Thus, lemma \ref{l322} is
false.
\end{proof}

\medskip In the next subsection we show a situation in which lemma
\ref{l322} does not hold. The construction resembles the $M_1$
structure of \citet*{muller06:_funny}.

\subsection{When the topological postulate is false}\label{imptop}

We will now show a situation in which lemma \ref{l322} does not hold.
By fixing two spatial dimensions we will restrict ourselves to
$\reals^2$, the first coordinate representing time.

As usual, $\Sigma$ is the set of all scenarios of a world $B$. Let $C$
be the set of all splitting points: \[ C := \bigcup_{\sigma,\eta \in
  \Sigma} C_{\sigma\eta}
\] We put
\begin{equation}\label{imptop1}
C := \{ \langle 0, n \rangle | n \in \mathbb{N} \cup \{ 0 \} \}
\end{equation}
The idea is that all splitting points are binary: any scenario passing
through a given splitting point can go either ``left'' or ``right''.
Since there are as many splitting points as natural numbers, we can
identify $\Sigma$ with a set of 01-sequences. Another requirement on
$\Sigma$ is that it contain only  sequences with finitely many 0s.
Let $G$ be a subset of $\Sigma$ containing only the sequence without
any 0s and all sequences that have all their 0s at the beginning. The
elements of $G$ will be labeled as below:
\begin{eqnarray*}\label{imptop2}
\sigma_0 = 1111..... \\ \sigma_1 = 01111.... \\ \sigma_2 = 00111....
\\ \sigma_3 = 00011....
\end{eqnarray*}
Let us next consider a sequence $Z_i^M$ of points in $\reals^2$ such
that for all $i \in \mathbb{N} \; z_i = \langle i-1/2,0 \rangle$. This
way, a given $z_i \in Z_i^M$ is in the Minkowskian sense above all
splitting points $\langle 0, n \rangle | n < i$ and above no other
splitting points.

Consider now a sequence $Z_i$ in $B$, $Z_i = \{ [z_i \sigma_i] | i \in
\mathbb{N} \}$. We will now show that $Z_i$ is a chain. Take any $[z_m
\sigma_m],[z_n \sigma_n] \in Z_i$ such that $m \neq n$. Either $m < n$
or $n < m$; suppose $m < n$ (the other case is analogous). Since $m <
n$, $z_m \leqslant_M z_n$. $z_m \in R_{\sigma_m \sigma_n}$ since it is
not above any splitting points between $\sigma_m$ and $\sigma_n$.
Therefore $z_m \sigma_m \equiv_S z_m \sigma_n$, so $[z_m \sigma_m]
\leqslant_S [z_n \sigma_n]$. We have shown that any two elements of
$Z_i$ are comparable by $\leqslant_S$. Therefore, $Z_i$ is a chain in
$B$, thus being an upward-directed subset of $B$.

The set of all upward-directed subsets of $B$ meets the requirements
of the Zorn-Kuratowski Lemma, since a set-theoretical sum of any chain
subset of it is also an upward-directed subset of $B$ and is an upper
bound of the chain with respect to inclusion. Therefore, there exists
a maximal upward-directed subset of $B$ (a history $h^*$) such that
$Z_i \subseteq h^*$. But lemma \ref{l322} is false with respect to
this history, since for all $\sigma \in \Sigma$, $h^* \neq \{
[x_\sigma]|x \in \reals^2 \}$! Suppose to the contrary, that for a
certain $\sigma \in \Sigma \; h = \{ [x_\sigma] | x \in \reals^2 \}$.
As a member of $\Sigma$, $\sigma$ has to contain a ``$1$'' at some
point $k$ (starting with $0$). Then both $[z_{k+1} \sigma_{k+1}] \in
h^*$ and $[z_{k+1} \sigma] \in h^*$, so $z_{k+1} \in R_{\sigma_k
  \sigma_{k+1}}$. But $C_{\sigma_k \sigma_{k+1}} \ni \langle 0, k
\rangle \leqslant_M z_{k+1}$, so $z_{k+1} \notin R_{\sigma_k
  \sigma_{k+1}}$ and thus we arrive at a contradiction.

We will now show that our topological postulate \ref{p1} is not met in
this situation. Consider a chain $Z := Z_i^M \cup \{ \langle -1,0
\rangle \}$. Note that $\langle -1,0 \rangle = inf(Z)$. Consider next
the chain topology on $\Sigma_{h^*}(\langle -1,0 \rangle)$ (as defined
in the last section) with $Z$ as the original chain. $\{
\Sigma_{h^*}(z_i) \}$ is a centred family of closed sets, but its
intersection is empty, as $\Sigma$ does not contain a scenario
corresponding to the sequence comprised of 0s only. Therefore, we
arrived at a contradiction with our corollary \ref{c1}, so postulate
\ref{p1} is not met: the chain topology is \textit{not} compact.

\section{Conclusions}\label{concl}
Having proved theorem \ref{th5}, we can adopt M\"{u}ller's
(\citeyear{muller_nato}) proof of the fact that $\langle B,
\leqslant_S \rangle$ meets all the requirements in definition
\ref{BSTmodel} and conclude that it is a model of BST. We keep in
mind, though, that we have introduced a new postulate \ref{p1} into
the proof and shown that it is not trivial (not always true). We will
demand that Minkowskian Branching Structures meet our topological
postulate. This way, a MBS is a special kind of BST model: its Our
World and ordering $\leqslant$ are constructed as, respectively, $B$
and $\leqslant_S$, as proposed by M\"{u}ller, and furthermore, our
postulate \ref{p1} is true in the model.

Due to the following self-evident Fact, we have fulfilled our promise
from the introduction and produced BST models in which histories are
isomorphic to Minkowski space-times.

\begin{fact}
  Let $\mathcal{W} = \langle W, \leqslant_S \rangle$ be an MBS and let
  $h$ be a history in $\emph{W}$ of the form $\{ [x_\sigma] \mid x \in
  \reals^4 \}$ for a certain $\sigma \in \Sigma$. Then
\begin{equation*}
  \langle h , \leqslant_S \mid_h \rangle \cong \langle \reals^4 , \leqslant_M \rangle
\end{equation*}
by means of the isomorphism $i: h \rightarrow \reals^4$ such that
$i([x_\sigma]) = x$.
\end{fact}

Let us observe that in contrast to general BST models, MBS models
involve numbers representing spatiotemporal locations, which raises
the question of whether the models are frame independent. The best way
to read an MBS model ${\mathcal W}$ is to consider it as produced by
an idealized omniscient observer associated with a given frame of
reference $F$. Of every possible event, the observer knows the event's
location (in his/her frame of reference), whether or not the event is
a choice point, and what possible continuations the event has.  There
might be another MBS model ${\mathcal W'}$ of the same world but
produced by another idealized omniscient observer associated with a
frame of reference $F'$. Since by the assumption, ${\mathcal W}$ and
${\mathcal W'}$ represent the same world, there is a one-to-one
correspondence $I$ between point events of the two models, with
$I([x_{\alpha}], [x'_{\alpha'}])$ read as `the same event described in
two different frames of reference'.  We then say that ${\mathcal W}$
and ${\mathcal W'}$ are relativistically invariant if for every
$[x_{\alpha}] \in W$ and $[x'_{\alpha'}] \in W'$ such that
$I([x_{\alpha}], [x'_{\alpha'}])$, $x'$ is the Lorentz transform of
$x$.  Thus, a distinction is looming between a single MBS model which
is frame-dependent, as it represents the world as seen from some frame
of reference, and the relativistically invariant set of MBS models
representing the world as seen from all (inertial) frames of
reference.

Finally, can we generalize the construction so as to have branching
structures produced out of space-times of general relativity?
Unfortunately, the answer is no. As already observed in
\citet{belnap92}, in general space-times of general relativity are not
upward directed.  That is, it is not (generally) true that for
incomparable $x, y$ from a given space-time, there is a $z$ such that
$x \leqslant z$ and $y \leqslant z$. Yet recall that in BST histories
must be upward-directed. Thus, the task of producing
generally-relativistic branching structures requires a major overhaul
of the BST theory.

\newpage
\section{Appendix: splitting points and choice points}\label{spcp}
Since it purports to establish that ``For histories $h_\sigma$,
$h_\eta$ $\subset B$ the set $C_{\sigma,\eta}$ is the set of choice
points'', Lemma 4 in \citet{muller_nato} seems to require
reformulation. A splitting point, as a member of $\reals^4$, is not a
member of $B$, and thus is not a choice point.

An obvious move would be to observe that every splitting point $x$ for
scenarios $\sigma$ and $\eta$ in a sense ``generates'' a choice point
for histories $h_\sigma$ and $h_\eta$. That is, if $x \in
C_{\sigma\eta}$ then $[x_\sigma]$ is maximal in $h_\sigma \cap
h_\eta$.

What might not be as evident is that, since we have dropped the
requirement of finitude of every $C_{\sigma\eta}$, the converse is not
true: in some cases there are choice points which are not
``generated'' in the above way by any splitting points. We will now
try to persuade the reader that this is indeed the case. The idea is
to use sequences of generated splitting points convergent to the same
point. The argument is simple in $\reals^2$, as we need only two
sequences, but it gets more complicated as the number of dimensions
increases. (For convenience, in the argument below we use the symbols
``$>_S$'' and ``$>_M$'' defined in the natural way and  based on
``$\leqslant_S$'' and ``$\leqslant_M$'', respectively.)

\begin{definition}
  1. $SC_{\sigma\eta}:=\{[c_\sigma]|c \in C_{\sigma\eta}\}$
\begin{eqnarray*}2. \textbf{C}_{\sigma\eta}:=\{[x_\gamma]:&(1)&
[x_\gamma] \in h_\sigma \cap h_\eta\mbox{~and}\\ &(2)& \forall z \in \reals^4 \forall \alpha \in \Sigma
([z_\alpha]
>_S [x_\gamma] \Rightarrow [z_\alpha] \notin h_\sigma \cap h_\eta)\
\end{eqnarray*}
\end{definition}

``$SC_{\sigma\eta}$'' is to be read as ``The set of generated choice
points for histories $h_\sigma$ and $h_\eta$''.

``$\textbf{C}_{\sigma\eta}$'' is to be read as ``The set of choice
points for histories $h_\sigma$ and $h_\eta$''.

It is of course irrelevant whether we choose $\sigma$ or $\eta$ in
square brackets in the definition of the set of generated choice
points, since if $c \in C_{\sigma\eta}$ then $c_\sigma \equiv_s
c_\eta$ and thus $[c_\sigma] = [c_\eta]$.

\begin{theorem}\label{LW1}{For some $C_{\sigma\eta}$, $SC_{\sigma\eta}
\varsubsetneq \textbf{C}_{\sigma\eta}$.}
\end{theorem}

\textit{Proof sketch.} Again, by fixing two spatial dimensions we will
restrict ourselves to $\reals^2$. Let $x=(0,0)$. Let $C_1=\{(0,1/n)|n
\in \mathbb{N} \backslash \{0\}\}$ and $C_2=\{(0,-1/n)|n \in
\mathbb{N} \backslash \{0\}\}$. Let $C_{\sigma\eta} = C_1 \cup C_2$.
As $x \notin C_{\sigma\eta}$, it is evident that $[x_\sigma] \notin
SC_{\sigma\eta}$. We will show that $[x_\sigma] \in
\textbf{C}_{\sigma\eta}$, thus proving the theorem.

We have to show that $[x_\sigma]$ meets conditions $(1)$ and $(2)$
from the above definition. As for $(1)$, $ \forall c \in
C_{\sigma\eta} \mbox{ x SLR c }$, so $x \in R_{\sigma\eta}$. It
follows that $x_\sigma \equiv_S x_\eta$ and finally (as it is obvious
that $[x_\sigma] \in h_\sigma$) that $[x_\sigma] \in h_\sigma \cap
h_\eta$.

Now for (2). Consider $[z_\alpha]$ such that (a) $[z_\alpha] >_S
[x_\sigma]$. By definition of $>_S$, $z >_M x$ and $x_\alpha =
x_\sigma$. Let $z=(z_0,z_1)$ (the first coordinate is temporal). We
distinguish two cases: either the spatial coordinate $z_1$ is equal to
$0$ or it is something else.

If $z=(z_0,0)$, take $k \in \reals$, $k<z_0$ such that $(0,k) \in
C_{\sigma\eta}$ (such a $k$ exists since $C_1$ converges to $(0,0)$).
(*) Since $D^2_M(z,(0,k)) = k - z_1 <0$, it follows that $x >_M (0,k)
\in C_{\sigma\eta}$.

On the other hand, if $z_1 \neq 0$, consider $v$ defined as follows:
\[
v :=\left\{\begin{array}{lll}1 & \textrm{if $z_1 \geq 1$}\\ z_1 &
    \textrm{if $z_1 \in (0,1) \cup (-1,0)$}\\-1 & \textrm{if $z_1
      \leqslant -1$}\end{array}\right.
\]

We choose $(0,k) \in C_{\sigma\eta}$ such that $0<k \leqslant v$ (if
$v$ is positive) or $v \leqslant k<0$ (if $v$ is negative). It is
always possible to find such a point, since both $C_1$ and $C_2$
converge to $(0,0)$. We have to prove that (b) $z >_M (0,k)$.

From (a) we know that (c) $z >_M (0,0)$. To arrive at (b) it suffices
to show that (d) $z >_M (0,v)$. From (c) it follows that (e) $z_0 \geq
z_1$. We have two cases to consider. First, if (f) $z_1 \geq 1$ or
$z_1 \leqslant -1$, $D^2_M(z,(0,v)) = -z_0^2 + (z_1 - 1)^2 = -z_0^2 +
z_1^2 + 1 - 2z_1$, which (by (f) and (e)) is below $0$, which fact is
equivalent to (d). Second, if $z_1 \in (0,1) \cup (-1,0)$,
$D^2_M(z,(0,v)) = -z_0^2 + (z_1 - z_1)^2 = -z_0^2$ which is of course
negative, so again we arrive at (d).

From (c) and (d) and from the requirement on choosing $(0,k)$ we get
the needed result (b).

Since $z >_M (0,k) \in C_{\sigma\eta}$, it is true that $z \notin
R_{\sigma\eta}$ and thus $[z_\alpha] \notin h_\sigma \cap h_\eta$. We
have thus proved that $[x_\sigma]$ fulfills condition (2).

Unfortunately already in $\reals^3$ the construction fails at point
(*). To overcome the problem we would have to use four sequences of
splitting points convergent to $(0,0,0)$ (intuitively situated at the
arms of the coordinate system). To deal with the situation in
$\reals^4$ we would have to similarly introduce six sequences
convergent to $(0,0,0,0)$. We will not dwell on the details here, as
the point being made does not seem to be significant enough in
proportion to the arduous complexity of the argument.

\begin{conjecture}
  For any scenarios $\sigma, \eta \in \Sigma$, the set
  $\textbf{C}_{\sigma\eta}$ contains exclusively points which belong
  to $SC_{\sigma\eta}$ or points $[x_\alpha]$ such that $x$ is a limit
  of a sequence of points belonging to $C_{\sigma\eta}$.
\end{conjecture}


\end{document}